\documentstyle[12pt,twoside]{article}
\begin{document}
\begin{center}

{\Large\bf ON THE CONSISTENCY OF A\\[5PT]
REPULSIVE GRAVITY PHASE\\[5PT]
IN THE EARLY UNIVERSE\\[5PT]}
\medskip
 
{\bf A.B. Batista\footnote{e-mail: brasil@cce.ufes.br},
J. C. Fabris\footnote{e-mail: fabris@cce.ufes.br} and S.V.B. Gon\c{c}alves\footnote{e-mail: sergio@cce.ufes.br}}  \medskip

Departamento de F\'{\i}sica, Universidade Federal do Esp\'{\i}rito Santo, 
29060-900, Vit\'oria, Esp\'{\i}rito Santo, Brazil \medskip\\
\medskip

\end{center}
 
\begin{abstract}
We exploit the possibility of existence of
a repulsive gravity phase in the evolution of the Universe.
A toy model with a
free scalar field minimally coupled to gravity, but with the "wrong sign" for the energy and negative
curvature for the spatial section, is studied in detail. The background solutions display a bouncing, non-singular
Universe. The model is well-behaved with respect to tensor perturbations. But, it exhibits growing models with
respect to scalar perturbations whose maximum occurs in the bouncing. Hence, large inhomogeneties are
produced. At least for this case, a repulsive phase may destroy homogeneity, and in this sense it may
be unstable. A newtonian analogous model is worked out; it displays
qualitatively the
same behaviour.
The generality of this result is discussed. In particular,
it is shown that the addition of an attractive radiative fluid
does not change essentially the results. We discuss also
a quantum version of the
classical repulsive phase, through the  Wheeler-de Witt equation in mini-superspace, and we show that it displays
essentially the
same scenario as the corresponding attractive phase.
\vspace{0.7cm}

PACS number(s): 04.20.Cv., 04.20.Me
\end{abstract}
 
\section{Introduction}

Gravity is an attractive interaction. This fact is represented by an universal positive coupling constant.
All avaliable observations confirm this statement. However, there is in principle
no a priori theoretical reason
to exclude the existence of a repulsive gravity interaction represented by a negative coupling
constant or, equivalently, a negative gravitational charge (mass). Indeed, repulsive cosmic interactions
have been proposed from time to time in the literature (see for example \cite{novello, mannheim}).
The question we would like to answer is the following: is a repulsive cosmic interaction physically acceptable?
To our point of view a specific type of interaction may
be considered as non-physical only if it is mathematically inconsistent and/or it leads to some
undesirable features like some kind of instability or divergence
in physical quantitities.
\par
The aim of this work is to verify to which extent a repulsive gravity may be considered as non-physical.
In \cite{novello} repulsive interactions in the primordial Universe appears as consequence of coupling
fermionic and electromagnetic field with gravity, which is coupled conformally to a scalar field.
A repulsive cosmic phase appears initially becoming attractive in a later phase, with
a changing of sign of the gravitational coupling.
In \cite{mannheim}, the repulsive gravity appears as consequence of the employment of conformal gravity,
through the Weyl's tensor: the conformal coupling to matter fields leads to a repulsive
gravity at cosmological level, keeping its attractive character at local level.
The curvature of the spatial section is necessarily negative. It has also been argued that the recent results concerning an accelerated expansion
of the Universe \cite{perlmutter,riess} are due to a cosmic repulsive effect \cite{mannheim1}. In fact, such
repulsive effect appears for example in the Brans-Dicke dust cosmological solution \cite{rose,gurevich} for negative values of
$\omega$, leading to a negative effective gravitational coupling.
In this case it is not clear how to reconcile this "cosmic" results with the local tests of gravity.
\par
In a matter of fact, repulsive gravity may appear in very different
contexts. The anti-de Sitter space-time, which is very popular today
due to the AdS/CFT correspondence \cite{klebanov}, is an example.
Anti-de Sitter space-time implies negative curvature in the spatial section.
Repulsive gravitational effects appears also when the conformal invariance
of a gravity-scalar field system
is broken. Let us consider for example the lagrangian
\begin{equation}
\label{l1}
L = \sqrt{-g}\biggr[\biggr(\frac{\phi^2}{6} - \frac{1}{3l^2}\biggl)R
+ \phi_{;\rho}\phi^{;\rho}\biggl] \quad ,
\end{equation}
where $l$ is a constant. The gravitational coupling is given
by $G \propto \biggr[\frac{\phi^2}{6} - \frac{1}{3l^2}\biggl]^{-1}$
and can become negative. This system can
be exactly solved when the curvature of the spatial section is negative;
the scale factor displays a non-singular behaviour and
the gravitational coupling is also negative \cite{orlov,gustavo}.  The system defined
by (\ref{l1}) differs from the spontaneously broken conformal symmetry of
\cite{mannheim} since the latter one is based on the Weyl tensor, instead of
a conformally coupled scalar field.
\par
On the other hand, it is generally argued that the initial singularity
may be avoided through the apparence of repulsive effects of quantum
origin. Indeed, quantum models in mini-superspace where there is, besides
gravity, a perfect fluid are examples of non-singular models obtained
through such effects. 
A carefull analysis of this problem,
performed in \cite{flavio1,nelson3}, indicates that a kind of
repulsive gravity effect appears as the Universe approaches the singularity.
\par
In fact, repulsive effects are frequently explicitly or implicitly refered to, mainly when comes to scene
situations where a singularity is presented, like in the big-bang scenario,
and one looks for a mechanism to avoid it. In principle, the violation of the strong energy condition seems
enough to do the job. But, we can think, in a more general level, that
repulsive gravity, violating all energy
conditions, can exist and become
the dominant interaction at some moment, avoiding the appearence of a singularity\footnote{Violation of all energy conditions occurs in many
situations in physics. Examples are the Casimir effect and the evaporation
of black holes\cite{visser}. Classically,
a self-interacting scalar field may also lead to violation of
the energy conditions. Our point of view
is that the energy conditions establish requirements to obtain certain physical effects, and
their violation does not imply the violation of any known physical law.}. This could be
the case if there is some repulsive gravity effect, represented by an universal negative coupling constant,
which dominates in extreme situations as it happens in
the very early Universe, becoming negligible later.
\par
We intend here to verify if such repulsive phase may lead to coherent cosmological scenario.
To perform this study, we will require this possible repulsive phase to be mathematically consistent
and not contain any anomaly either at classical and quantum level. To be explicit, we will work out
a toy model consisting of Einstein's equations coupled minimally to a free scalar field. The possibility that
this scalar field appears in Einstein's equation with a wrong sign will be exploited. In spite of the simplicity of this model, it can be connected
to some of the
above quoted anti-gravity examples through some suitable transformations.
It will be shown that
a consistent and singularity-free background cosmological scenario comes out if the curvature of the spatial section is
negative. Another advantage of this simple toy model is that it does not require
the introduction of other kind of matter field: a single fluid repulsive
model suffices. But, we will also discuss the coupling of this
repulsive fluid with ordinary matter, what constitute a more realistic
situation for later phases in the evolution of the Universe.
\par
In some sense, this "exotic" scalar field ameliorates things with respect to the "normal" scalar field,
which leads to a singular cosmological scenario. But, a perturbative analysis will reveal the presence of some kind of instabilities
near the bounce, exactly where the repulsive phase becomes dominant: the Universe may become too inhomogeneous there,
and the initial hypothesis of isotropy and homogeneity may not survive during the bouncing. This results are
somehow supported by a similar newtonian analysis. The introduction of
an attractive radiative fluid is not able to surmount these instability
problems
\par
Later, we will verify if the some more anomaly of this "exotic" field can appear in its quantum behaviour.
This is a very difficult question to answer, since there is not yet a full consistent quantum gravity theory and
the study of quantum effects in a fixed background represents a very stringent simplification of our dynamical
model. However, we try to obtain some hints about a posssible answer to that question studying the Wheeler-de Witt
equation for that model in mini-superspace. In this case, the equations are exactly solvable. We verify that the
solutions with negative and positive energy at this quantum level are essentially the same, and no further
restriction can be obtained through this way for a repulsive gravity phase,
unless some specifications concerning boundary conditions are made.
\par
In the next section we describe our toy model, determining the background solutions. In section $3$
a perturbative analysis of the previously found solution is carried out. In section $4$ we couple the repulsive fluid to an attractive one, in the
present
case a radiative fluid, and we show that the instabilities remain. In
section $5$ the
Wheeler-de Witt equation in mini-superspace is solved both for the "normal" and "exotic" scalar field, and
the results are compared. In section $6$ the conclusions are presented.

\section{A scalar field model}

Let us consider gravity coupled to a scalar field, but with no restriction to the sign of this coupling.
Hence, in our conventions \footnote{The conventions are $R_{\mu\nu} = \partial_\rho{\Gamma^\rho}_{\mu\nu} -
\partial_\nu{\Gamma^\rho}_{\mu\rho} + ...$, sign$g = (+ - - -)$.}, the lagrangian reads
\begin{equation}
\label{lag}
L = \sqrt{-g}\biggr\{R \pm \biggr(\phi_{;\rho}\phi^{;\rho} - 2V(\phi)\biggl)\biggl\}
\end{equation}
where the positive (negative) sign inside the brackets implies negative (positive) energy density, hence
repulsive (attractive) gravitational effects.
From this lagrangian, it follows the field equations,
\begin{eqnarray}
\label{fe1}
R_{\mu\nu} - \frac{1}{2}g_{\mu\nu}R &=& \mp \biggr\{\phi_{;\mu}\phi_{;\nu} - \frac{1}{2}g_{\mu\nu}\phi_{;\rho}\phi^{;\rho} + g_{\mu\nu}V(\phi)\biggl\} \quad , \\
\label{fe2}
\Box\phi &=& - V_\phi(\phi) \quad .
\end{eqnarray}
An anti-gravity phase is possible both in the self interacting as in the
free scalar field models.
However, we will neglect the possible self-interaction of the scalar field
in order to obtain analytical expressions. The resulting
free scalar field model, in spite of its simplicity, may cover many different situations quoted before.
For example, the theory described by (\ref{l1}) can be recast in the
form (\ref{lag}), with $V(\phi) = 0$, through a field redefinition and a
conformal transformation.
\par
The geometry is described by the Friedmann-Robertson-Walker metric,
\begin{equation}
\label{m}
ds^2 = dt^2 - a^2(t)\biggr[\frac{dr^2}{1 - kr^2} + r^2(d\theta^2 + \sin^2\theta d\phi^2)\biggl]
\end{equation}
where $k = 0, 1, -1$ depending if the spatial section is flat, closed or opened.
Since we will be interested in a homogeneous and isotropic Universe, the scalar field
$\phi$ must be function only of the cosmic time: $\phi = \phi(t)$.
\par
Inserting the metric (\ref{m}) into the field equations (\ref{fe1},\ref{fe2}), fixing $V(\phi) = 0$, we obtain
the equations of motion,
\begin{eqnarray}
\label{em1}
3\biggr(\frac{\dot a}{a}\biggl)^2 + 3\frac{k}{a^2}
&=& \mp \frac{1}{2}\dot\phi^2
\quad , \\
\label{em2}
\ddot\phi + 3\frac{\dot a}{a}\dot\phi &=& 0 \quad .
\end{eqnarray}
When we choose the upper sign in (\ref{em1}) there is no real solution for $a(t)$ unless $k = - 1$. Hence, we
will consider just the open Universe. Anyway, negative curvature is
a general feature of the cases quoted before. From (\ref{em2}), we obtain $\dot\phi = \phi_0a^{-3}$. Inserting this
first integral in (\ref{em1}), and changing to the conformal time $dt = a(\eta)d\eta$, we find the following
solutions:
\begin{itemize}
\item Upper sign:
\begin{eqnarray}
\label{bs1a}
a(\eta) &=&  a_0\sqrt{\cosh 2\eta} \quad ,\\
\label{bs1b}
\phi(\eta) &=& \phi_0\arctan \biggr[(\cosh\eta \pm \sinh\eta)^2\biggl] \quad ;
\end{eqnarray}
\item Lower sign:
\begin{eqnarray}
\label{bs2a}
a(\eta) &=& a_0\sqrt{\sinh 2\eta} \quad , \\
\label{bs2b}
\phi(\eta) &=& \phi_0\ln\tanh\eta \quad .
\end{eqnarray}
\end{itemize}
\par
The solutions corresponding to a repulsive gravity phase (upper sign) describe a non-singular
Universe. The positive energy solutions (lower sign)
describe an Universe with an initial singularity. Hence, from the point of view of the
background, the negative energy solutions seem more interesting than the positive energy solutions,
due to the existence of an initial singularity in the latter.
We can ask if such
bouncing, eternal, singularity-free Universe is not plagued with instability problems.
This is the subject of the next section.
\par
Before turn to this perturbative analysis, one may ask if the existence of a repulsive phase is restricted to open geometries. In fact, if we consider just
one field, this is true. However, in mixed models, where there is a fluid
which acts attractivelly and other repulsivelly,
is it possible to have solutions even if the curvature is non-negative?
Let us consider for example a flat Universe with a radiative
fluid whose gravitational
coupling is positive and a stiff matter fluid (or, equivalently,
a free scalar field)
whose gravitational coupling is negative. Both fluids interact
between themselves only through the geometry.
The equations of motion are reduced essentially to
\begin{equation}
\biggr(\frac{\dot a}{a}\biggl)^2 = \frac{k_1}{a^4} - \frac{k_2}{a^6} \quad ,
\end{equation}
where $k_1$ and $k_2$ are constants. The spatial curvature is zero. The solution for this equation, in terms of
the conformal time, is
\begin{equation}
a = a_0\sqrt{\eta^2 + C^2} \quad ,
\end{equation}
where $C$ is a constant. This solution describes a bounce Universe, which goes asymptotically to a radiation dominated Universe.
It is curious that if the roles of these fields were interchanged, leading to a repulsive radiation field and an
attractive stiff matter field, the solution
should be written as
\begin{equation}
a = a_0\sqrt{C^2 - \eta^2} \quad .
\end{equation}
This is a singular Universe which can exist only for a finite period of time.

\section{Evolution of tensor and scalar perturbations}

We will now perturb the scalar field model described before.
Hence, we write the metric and the scalar field as
$g_{\mu\nu} = {\stackrel{0}{g}}_{\mu\nu} + h_{\mu\nu}$ and $\phi = \stackrel{0}{\phi} + \delta\phi$,
where ${\stackrel{0}{g}}_{\mu\nu}$ and $\stackrel{0}{\phi}$ are the background solutions found
previously; $h_{\mu\nu}$ and $\delta\phi$ are small perturbations around them.
Due to the coordinate transformation invariance of the field equations, we can
impose a coordinate condition. We choose the synchronous coordinate condition $h_{\mu0} = 0$; the only
non-negative components will be those such that $\mu = i$ and $\nu = j$.
The perturbations can be decomposed in purelly tensor perturbations, vectorial perturbations and scalar perturbations \cite{weinberg}.
The vectorial perturbations are quite trivial, leading to terms that are proportional to the inverse of the scale factor.
Tensor and scalar perturbations are less trivial, and we study them separately.
\par
In order to perform this perturbative analysis, we
write the field equations as
\begin{eqnarray}
\label{nfe1}
R_{\mu\nu} &=& \mp \phi_{;\mu}\phi_{;\nu} \quad , \\
\label{nfe2}
\Box\phi &=& 0 \quad .
\end{eqnarray}

\subsection{Tensor perturbations}

Perturbing (\ref{nfe1}), retaining only the transverse tracelless component, the metric perturbations are described by a single equation
which reads, in terms of the conformal time,
\begin{equation}
\label{gwe}
h'' - 2\frac{a'}{a}h' + \biggr\{\tilde n^2 - 2 \biggr(\frac{a''}{a} - \frac{a'^2}{a^2}\biggl)\biggl\}h = 0 \quad .
\end{equation}
To obtain this equation the metric perturbation has been written as $h_{ij}(\eta,\vec x) = h(\eta)Q_{ij}(\vec x)$,
such that $Q_{kk} = Q_{ik|k} = 0$ and $\nabla^2Q_{ij} = - n^2Q_{ij}$. Hence, the $Q_{ij}$ are eigenfunctions
of the laplacian operator defined in the constant curvature spatial hypersurface. Moreover $\tilde n^2 = n^2 + 2k =
n^2 - 2$, due to the fact that the curvature is negative.
\par
We must insert in (\ref{gwe}) the solutions (\ref{bs1a}) and (\ref{bs2a}). For the first case, which corresponds to
the repulsive scalar field, we find the equation
\begin{equation}
h'' - 2\tanh2\eta h' + \biggr\{\tilde n^2 - \frac{4}{\cosh^22\eta}\biggl\}h = 0 \quad .
\end{equation}
This equation may be solved through the transformations $h(\eta) = \lambda(\eta)\cosh2\eta$ and $x = \sinh2\eta$ by which
it takes the form
\begin{equation}
(1 + x^2)\ddot\lambda + 2x\dot\lambda + \frac{\tilde n^2}{4}\lambda = 0 \quad,
\end{equation}
where dot means derivative with respect to $x$. This equation admits a solution under the form of hypergeometric
functions. The final expressions are:
\begin{eqnarray}
h_1(\eta) &=& \cosh2\eta\,{ _2F_1}\biggr(\frac{1}{4}(1 + \sqrt{1 - \tilde n^2}),\frac{1}{4}(1 - \sqrt{1 - \tilde n^2}), \frac{1}{2},
- \sinh^2\eta\biggl) \quad , \\
h_2(\eta) &=& \sinh2\eta\cosh2\eta\,{ _2F_1}\biggr(\frac{1}{4}(3 + \sqrt{1 - \tilde n^2}),\frac{1}{4}(3 - \sqrt{1 - \tilde n^2}), \frac{3}{2},
- \sinh^2\eta\biggl) \quad .
\end{eqnarray}
From the previous definitions we have $- 2 \leq \tilde n^2 < \infty$.
\par
These solutions for the tensor modes exhibit a quite regular behaviour. The perturbations tends to diverge at both
asymptotics, where anyway the scale factor goes to infinity; but they are strongly supressed as they approach the
bounce. The general features are sketched in figures 1 and 2. Hence, the background solutions are stable against tensor perturbations: the production of gravitational waves
does not destroy the configuration. This result will be discussed later.
\par
Repeating the same analysis for the positive energy solution, we find the following expressions for the
tensor perturbations in this case:
\begin{eqnarray}
h_3(\eta) &=& \sinh2\eta\,{ _2F_1}\biggr(\frac{1}{4}(1 + \sqrt{1 - \tilde n^2}),\frac{1}{4}(1 - \sqrt{1 - \tilde n^2}), \frac{1}{2},
\cosh^2\eta\biggl) \quad , \\
h_4(\eta) &=& \sinh2\eta\cosh2\eta\,{ _2F_1}\biggr(\frac{1}{4}(3 + \sqrt{1 - \tilde n^2}),\frac{1}{4}(3 - \sqrt{1 - \tilde n^2}), \frac{3}{2},\cosh^2\eta\biggl) \quad .
\end{eqnarray}
The behaviour of these solutions are displayed in figures 3 and 4. They have essentially the same features of the
previous case, but "cutting" the half of the graph.  No anomalous behaviour occurs, and all perturbations emerge
from a zero value from the singularity, exhibiting growing or oscillatory behaviour, depending on the value
of $n$, at the infinity asymptotic.

\subsection{Scalar perturbations}

Now, we turn to the scalar modes.
We decompose the metric as $h_{ij}(\eta,\vec x) = h_{ij}(\eta)Q(\vec x)$, where $\nabla^2Q = - n^2Q$.
Perturbing (\ref{nfe1},\ref{nfe2}), we obtain the following equations:
\begin{eqnarray}
\label{spe1}
h'' + \frac{a'}{a}h' &=& \mp 4\phi'\delta\phi' \quad , \\
\label{spe2}
\delta\phi'' + 2\frac{a'}{a}\delta\phi' + n^2\delta\phi &=& \frac{1}{2}\phi'h' \quad ,
\end{eqnarray}
where $h = \frac{h_{kk}}{a^2}$.
From (\ref{spe2}), we can express $h'$ and $h''$ in terms of $\delta\phi$ and its derivatives.
We obtain a third order differential equation:
\begin{equation}
\delta\phi''' + \biggr(3\frac{a'}{a} - \frac{\phi''}{\phi'}\biggl)\delta\phi' +
\biggr(n^2 + 2\frac{a''}{a} - 12\frac{a'^2}{a^2} + 12 - 2\frac{a'}{a}\frac{\phi''}{\phi'}\biggl)\delta\phi' +
\biggr(\frac{a'}{a} - \frac{\phi''}{\phi'}\biggl)n^2\delta\phi = 0 \quad .
\end{equation}
\par
The fact that we end up with a third order equation is a consequence of the coordinate condition chosen: the
synchronous coordinate condition contains a residual coordinate freedom, as it is well known \cite{peebles}.
This enables us to reduce the order of the equation, since we know the effect on $\delta\phi$ of this residual
coordinate freedom.
\par
First we will consider the negative energy case. Inserting the background solutions and defining
the transformation
\begin{equation}
\delta\phi(\eta) = \lambda(\eta)\cosh^{-3/2}2\eta \quad ,
\end{equation}
we obtain the equation
\begin{equation}
\lambda''' - 4\tanh2\eta\lambda'' + \biggr(n^2 + 5\tanh^22\eta - 2\biggl)\lambda' = 0
\end{equation}
which can be solved through the same kind of transformations as in the case of the tensor perturbations.
The final solutions are:
\begin{eqnarray}
\delta\phi_1(\eta) &=& \cosh^{-3/2}2\eta\biggr[C_1\int\cosh^{1/2}2\eta\times\nonumber\\
& &{_2F_1}\biggr(-1/4(1 + \sqrt{1 - n^2}\biggl),
- \frac{1}{4}(1 - \sqrt{1 - n^2}), \frac{1}{2}, - \sinh^22\eta\biggl) + C_2\biggl]\quad , \\
\delta\phi_2(\eta) &=& \cosh^{-3/2}2\eta\biggr[C_3\int\cosh^{1/2}\eta\sinh2\eta\times\nonumber\\
& & {_2F_1}\biggr(-1/4(1 + \sqrt{3 - n^2}\biggl),
- \frac{1}{4}(3 - \sqrt{1 - n^2}), \frac{3}{2}, - \sinh^22\eta\biggl) + C_4\biggl] \quad .
\end{eqnarray}
The $C_i$ are integration constants.
\par
For the positive energy case, the procedure is essentially the same, and the final expressions are
\begin{eqnarray}
\delta\phi_3(\eta) &=& \sinh^{-3/2}2\eta\biggr[D_1\int\sinh^{1/2}(\eta)\times\nonumber\\
& &{_2F_1}\biggr(-1/4(1 + \sqrt{1 - n^2}\biggl),
- \frac{1}{4}(1 - \sqrt{1 - n^2}), \frac{1}{2}, \cosh^22\eta\biggl) + D_2\biggl] \quad , \\
\delta\phi_4(\eta) &=& \sinh^{-3/2}2\eta\biggr[D_3\int\cosh2\eta\sinh^{1/2}2\eta\times\nonumber\\
& &{_2F_1}\biggr(-1/4(1 + \sqrt{3 - n^2}\biggl),
- \frac{1}{4}(3 - \sqrt{1 - n^2}), \frac{3}{2}, \cosh^22\eta\biggl) + D_4\biggl] \quad .
\end{eqnarray}
Again, the $D_i$ are integration constants.
\par
The analysis of the results obtained above are more involved. In principle we are tempted to
say that the perturbations are regular, since the integrand are regular. They diverge
at infinity; this is not a serious divergence. But, for the case of repulsive coupling,
the integration leads to very large values of the perturbations near the bounce.
The earlier the perturbations are originated, the larger are their values near
the bounce. For perturbations originated in the first asymptotic, their amplitude in the bouncing
tends to a divergent value.
That means that the Universe may become highly inhomogeneous near the bounce, or even unstable if we allow the perturbations
to orginate at $\eta \rightarrow - \infty$, and the
initial configuration is destroyed. This result
has been confirmed by direct numerical integration.
On the other hand, the positive energy solutions
display a more regular behaviour throughout the evolution of the Universe, as it is shown in figure $6$.
\par
It could be argued that the relevant quantity is not $\delta\phi$, but the associated density contrast,
$\Delta_\phi = \frac{\delta\phi'}{\phi'}$. However, for the negative energy case, we verified that this quantity takes also very large
values near the bounce, exhibiting essentially the same behaviour as $\delta\phi$.

\subsection{A newtonian analysis}

The preceding result can be confirmed qualitatively through a newtonian model.
This model is described by the following equations \cite{weinberg}:
\begin{eqnarray}
\label{bne1}
\dot\rho + \nabla.(\rho\vec v) &=& 0 \quad , \\
\label{bne2}
\dot{\vec v} + \vec v.\nabla(\vec v) &=& - \frac{\nabla p}{\rho} + \vec g \quad , \\
\label{bne3}
\nabla.\vec g &=& 4\pi G\rho \quad ,
\end{eqnarray}
where we have considered the "wrong" sign for the gravitational coupling.
These equations admit solutions for a isotropic and homogeneous Universe such that
\begin{equation}
\rho = \frac{\rho_0}{a^3} \quad , \quad \vec v = \vec r\frac{\dot a}{a} \quad , \quad \vec g =  \vec r\frac{4\pi G\rho}{3} \quad ,
\end{equation}
where the function $a$ obeys the equations
\begin{equation}
\biggr(\frac{\dot a}{a}\biggl)^2 + \frac{k}{a^2} = - \frac{8\pi G\rho}{3} \quad , \quad \frac{\ddot a}{a} = \frac{4\pi G\rho}{3} \quad .
\end{equation}
There is a solution if $k < 0$. The solution, in terms of the conformal
time, is
\begin{equation}
a(\eta) = a_0(1 + \cosh\eta) \quad ,
\end{equation}
which describes also a bouncing Universe. 
\par
Perturbing the equations (\ref{bne1},\ref{bne2},\ref{bne3}), we can determine
an equation for the density contrast
$\Delta = \frac{\delta\rho}{\rho}$, which reads:
\begin{equation}
\label{pne}
\Delta'' + \frac{a'}{a}\Delta' + \biggr\{n^2v_s^2 + 3\biggr(\frac{a''}{a} - \frac{a'^2}{a^2}\biggl)\biggl\}\Delta = 0 
\end{equation}
where $n$ denotes, as before, the wavenumber of the perturbation. This equation differs from the attractive gravity case
by the form of the function $a(\eta)$ and by the sign of the last term in (\ref{pne}). There is analytical solutions
for (\ref{pne}) in the long wavelength limit ($n \rightarrow 0$):
\begin{eqnarray}
\Delta_+ &=& E_1\biggr( - \frac{3\eta\sinh\eta}{(1 + \cosh\eta)^2} + \frac{5 - \cosh\eta}{1 + \cosh\eta}\biggl) \quad ;\\
\Delta_- &=& E_2\frac{\sinh\eta}{(1 + \cosh\eta)^2} \quad . 
\end{eqnarray}
$E_1$ and $E_2$ are integration constants. It is easily to see that the "growing mode" represented by the solution
$\Delta_+$ has a maximum near the bouncing. Qualitatively, the relavistic result is reproduced. However, the
maximum in the newtonian case is not so important as the relativistic case. This seems to be due to the contribution
of the pressure to the effective mass in the relativistic situation,
increasing the repulsive effect.

\section{Coupling with ordinary matter}

The repulsive gravity model developped in the last section can exist for
negative curvature of the spatial section and it is unstable. However,
matter must be important from some moment in the evolution of the Universe
on.
The introduction of ordinary (attractive) matter stabilizes the configuration?
Let us consider now this possibility.
\par
In the primordial Universe, a radiative fluid must dominate over the
dust fluid. Hence, it will be analyzed a two fluid model with radiation
and the repulsive scalar field. Only the negative spatial section
will be considered. Since a free scalar
field is equivalent to a stiff matter fluid, we will represent it by
$\rho_S$, with $p_S = \rho_S$. The Einstein equations reduce to
\begin{equation}
\frac{\dot a^2}{a^2} - \frac{1}{a^2} = \frac{8\pi G}{3}\biggr(\frac{\rho_1}{a^4}
-\frac{\rho_2}{a^6}\biggl) \quad ,
\end{equation}
where $\rho_1$ and $\rho_2$ are integration constants.
This equation admits the solution
\begin{equation}
a = a_0\biggr\{\biggr(\sqrt{\frac{4c_1}{c_2^2} + 1}\biggl)\cosh2\eta - 1\biggl\}^{1/2}
\end{equation}
with $c_1 = \frac{8\pi G\rho1}{3}$ and $c_2 = \frac{8\pi G\rho_2}{3}$.
Again, the scale factor behaviour reveals a non-singular, eternal
Universe. The presence of the stiff matter repulsive fluid still leads to
the avoidance of the singularity even in the presence of an attractive
radiative fluid.
\par
The perturbed equations can be obtained in the usual way. In a two
fluid model, where one of them is repulsive, the perturbed equations
take the form
\begin{eqnarray}
\ddot h + 2\frac{\dot a}{a}\dot h =\frac{1}{\alpha_M - \alpha_S}
\biggr[- (1 + 3\alpha_M)\biggr(2\frac{\ddot a}{a} + (1 + 3\alpha_S)
\frac{\dot a^2}{a^2} + \frac{k}{a^2}\biggl)\Delta_M & & \nonumber \\
+ (1 + 3\alpha_S)\biggr(2\frac{\ddot a}{a} + (1 + 3\alpha_M)
\frac{\dot a^2}{a^2} + 2\frac{k}{a^2}\biggl)\Delta_S\biggl] & &\, ,\\
\nonumber\\
\dot\Delta_M + (1 + \alpha_M)\biggr(\Psi - \frac{\dot h}{2}\biggl) &=& 0\, ,\\
\nonumber\\
(1 + \alpha_M)\biggr[\dot\Psi + (2 - 3\alpha_M)\frac{\dot a}{a}\Psi\biggl]
- \frac{n^2}{a^2}\alpha_M\Delta_M &=& 0 \, ,\\
\nonumber \\
\dot\Delta_S + (1 + \alpha_S)\biggr(\theta - \frac{\dot h}{2}\biggl) &=& 0\, ,\\
\nonumber\\
(1 + \alpha_S)\biggr[\dot\theta + (2 - 3\alpha_S)\frac{\dot a}{a}\theta\biggl]
- \frac{n^2}{a^2}\alpha_S\Delta_S &=& 0 \, .
\end{eqnarray}
In these equations $\alpha_M$ and $\alpha_S$ stand for the equation of
state of ordinary and repulsive fluids respectively. $\Psi$ and $\theta$
are the perturbations in the velocity of the normal and repulsive fluids,
$\Delta_M$ and $\Delta_S$ are the density contrast for each fluid
while $h$ is defined as before.
Fixing $\alpha_M = \frac{1}{3}$ and $\alpha_S = 1$, changing to
the conformal time, and inserting the background solutions
we obtain a system of coupled differential equations for $h$, $\Psi$,
$\theta$, $\Delta_M$ and $\Delta_S$. The resulting system of equations
seems to admit no exact solution. But, it can be studied numerically.
Figure $7$ displays the result of this numerical integration
for the density contrast of the radiative fluid: the perturbations
diverge in a finite time. Hence, the instability remains in this
two fluid model: ordinary matter is not able to stabilize the
model.

\section{A quantum cosmological analysis}

One may ask if some
anomalies may also be present at quantum level. A complete answer to this problem seems
very difficult to obtain. We could study the presence of quantum fields in the background
described before; but since the background
is completly regular, we may guess that no anomolous behaviour can
be expected from this approach. Or we may turn to the Wheeler-de Witt equation \cite{wdw,nelson}. In this
case, the first simplification we are, in some sense, obliged to make is to freeze out
all degrees of freedoms except those related to the scale factor and the scalar field,
working in the so-called mini-superspace.
Perhaps, we could take more degrees of freedom into account, but we do not think that
the final results would be modified drastically.
\par
Hence, we begin from the lagrangian (\ref{lag}). Inserting in it the metric (\ref{m}) and
the time-dependence of the scalar field, we end up, after integration by parts, with the expression
\begin{equation}
L = \biggr\{6\frac{a\dot a^2}{\sqrt{N}} - 6ka\sqrt{N} \pm \frac{a^3\dot\phi^2}{\sqrt{N}}\biggl\} \quad ,
\end{equation}
where $N$ is the lapse function connected with the time reparametrization freedom.
From it, the conjugate momenta are obtained, through the relation $\phi_q = \frac{\partial L}{\partial\dot q}$:
\begin{equation}
\pi_a = 12\frac{a\dot a}{\sqrt{N}} \quad , \quad \pi_\phi = \pm 2\frac{a^3\dot\phi}{\sqrt{N}} \quad .
\end{equation}
Hence, the hamiltonian can be constructed canonically:
\begin{equation}
H \sqrt{N}\biggr\{\frac{1}{24}\frac{\pi_a^2}{a} \pm \frac{1}{4}\frac{\pi_\phi^2}{a^3} + 6ka\biggl\} \quad .
\end{equation}
We must remark that in the derivation of the above hamiltonian, we have discarded a surface term. Hence,
the manifold must be compact. It is possible to have compact manifolds even if the spatial curvature
is zero or negative, as we want to consider here.
\par
The canonical quantization is now applied. The hamiltonian is taken as an operator which acts over a
wavefunction. The substitutions $\pi_a \rightarrow - i\hbar\frac{\partial}{\partial_a}$ and
$\pi_\phi \rightarrow - i\hbar\frac{\partial}{\partial_\phi}$ are included. Then, the Wheeler-de Witt
equation in the mini-superspace is obtained (with $\hbar = 1$):
\begin{equation}
\biggr\{\partial_a^2 + \frac{p}{a}\partial_a \pm \frac{6}{a^2}\biggr[\partial_\phi^2 + \frac{q}{\phi}\biggl]
- 144ka^2\biggl\}\Psi = 0 \quad .
\end{equation}
The parameters $p$ and $q$ were introduced to take into account the order ambiguity. The upper (lower) sign
in the second term refers to a classical negative (positive) energy. 
The Wheeler-de Witt equation may be solved by the separation of variable method. Hence, we writte
$\Psi(a,\phi) = \alpha(a)\beta(\phi)$. Taking $k = - 1$, two equations are obtained:
\begin{eqnarray}
\label{wdw1}
\alpha'' + \frac{p}{a}\alpha' + \biggr\{144a^2 \pm 6\frac{c}{a^2}\biggl\}\alpha &=& 0 \quad , \\
\label{wdw2}
\beta'' + \frac{q}{\phi} - c\beta &=& 0 \quad ,
\end{eqnarray}
where $c$ is a constant connected to the separation of variables. An important aspect is that the possibility
of having negative (positive) energy implies to choose the upper (lower) sign in (\ref{wdw1}). Since $c$ may
have any value, the choice of one possibility or another changes nothing in the final
analysis. According to this simplified quantum model, positive or negative classical energies seem to lead essentially to
the same quantum model.
\par
This last statement must, however, be taken with care. Indeed, the solutions
of the Wheeler-de Witt equation must be complemented by the choice of boundary
conditions. The choice of boundary conditions is not an easy task, and
in many cases, like the present one,
it is an open subject. In some cases, it is possible to fix
the boundary conditions by imposing the hamiltonian operator to be
self-adjoint. If we consider a specific boundary condition, it is possible
to select some solutions, and employ some interpretation scheme
(bohmian trajectories, expectation value of the scale factor,
semi-classical approach) to obtain previsions on the behaviour of
the Universe. The choice of the upper or lower sign in (\ref{wdw1}) implies
to change from a hyperbolic Wheeler-de Witt equation to an elliptical one.
This must lead to different boundary conditions. Even if the
space of solutions are the same in both cases, different boundary conditions
will lead to different sub-spaces of solutions.
\par
In order to be complete, we write down the solutions for equations (\ref{wdw1},\ref{wdw2}). They read,
\begin{eqnarray}
\alpha &=& a^{(1-p)/2}\biggr\{A_1J_\nu(\sqrt{36}a^2) + A_2J_{-\nu}(\sqrt{36}a^2)\biggl\} \quad , \\
\beta &=& \phi^{(1-q)/2}\biggr\{B_1I_r(c\phi) + B_2K_r(c\phi)\biggl\} \quad ,
\end{eqnarray}
where $J$, $I$ and $K$ are Bessel's function and modified Bessel's function,
$\nu = \frac{1}{4}\sqrt{1 - p \pm 6c}$, $r = \frac{1}{2}(1 - q)$. $A_i$ and $B_i$ are integration constants.
\par
These solutions, in a slightly different form, were analyzed in \cite{nelson1,nelson2}. In general, they predict
a singular Universe. However, gaussian superposition of these solutions may lead to non-singular universe \cite{nelson2}. Hence, both classical cases typical of positive or negative energies are covered by these solutions
of the Wheeler-de Witt equation.

\section{Final comments}

In this paper, we worked out a cosmological model including a repulsive phase verifying if it is consistent physically and
mathematically. In order to be specific, a model containing a minimal coupling between gravity and a free scalar field
has been studied. In spite of its simplicity, this model may
be connected with more general ones through some transformations.
A repulsive phase is obtained when the energy of this scalar field is negative. Mathematically consistent
solutions can be obtained only if the spatial curvature is negative. This fact has already been remarked by
\cite{mannheim} in the context of a conformal gravity. We showed that repulsive effects may appear
in cosmology with $k = 0$ only when mixed with some other attractive fluid. From the physical point of view, we tried to verify if these solutions are stable and if they present
some anomalies at the level of the Wheeler-de Witt equation in the mini-superspace.
\par
The first feature to be noticed is that, while the positive energy solution
is singular, the negative energy one presents a bounce,
being non-singular. Surprisingly, this seems to favor the negative energy solution. We turned then to a perturbative
analysis for both cases. Scalar and tensor perturbations were studied. Tensor perturbations present a very regular behaviour, diverging
only in the asymptotical limit where the scale factor also goes to infinity. This is not dangerous of course.
But, scalar perturbations reveal exactly the opposite behaviour: they are regular in the asymptotics, but assume
extremelly large values near the bounce, where the scale factor reachs its minimum value.
At this moment, the repulsive effects are the dominant ones, and even if they lead to nice features for the
background, they carry also the seed of their destruction: the Universe becomes too inhomogeneous.
This results has been qualitatively confirmed through a newtonian analysis.
Moreover, even if we couple the repulsive fluid to a normal, attractive,
fluid (here represented by a radiative fluid), the instabilities remain.
\par
These different results for tensor and scalar perturbations are not so surprising. Tensor perturbations are
sensitive essentially to the scale factor behaviour and not directly
to the fluid content.
The regularity observed in the tensor perturbations seems just to indicate that it is possible to have stable bounce scenario, which
can be obtained with weaker hypothesis, like the violation of the strong energy condition only.
On the other hand, scalar perturbations are sensitive to the matter content and couplings of the model. Hence,
they feel directly the repulsive character of the scalar field introduced here:
the repulsive field may lead to a large amplification of perturbations in a finite time, when the scale factor
takes its minimum value, from where the destruction of the homogeneity. In this sense, the model is unstable.
\par
The quantum model developed here, with a minisuperspace formulation of the Wheeler-de Witt equation, did not reveal any
anomaly in the negative energy model. In fact, the solutions with positive energy are indistinguishable of those with
negative energy; perhaps the boundary conditions for each case
may be different and may lead to different selection
of those solutions.
Moreover, previous analysis made of the positive energy model \cite{nelson1,nelson2} showed
the existence of singular and non-singular models. Hence, both possibilities corresponding to the classical solutions are covered.
\par
The main conclusion of this work is that a repulsive phase may lead to a non-singular Universe but which is
unstable in the sense that the initial hypothesis, like homogeneity, can not survive the
repulsive era. One may ask how general is this result. Of course, we have studied a very specific model. But, this model is
somehow similar, from the point of view of the behaviour of the scale factor, to those exposed in \cite{mannheim,flavio1,nelson3}; hence, we can argue that
the cosmic repulsion coming from the conformal gravity may
also be unstable, in the sense employed here, even if the study of this question for the conformal gravity
theory is much more involved due to the complexity of field content. On the other hand, the model presented in
\cite{novello} indicates a changing of the sign of the cosmological constant, and it has already been shown in other
situations that such transition from anti-gravity to gravity phase leads to instabilities
\cite{starobinski,fabris}. Hence, even if a deeper and
more general study is needed, all these results suggest the instability
of a cosmic repulsion phase, at least in models where the repulsive
phase is allowed due to a negative curvature of the spatial section.
\newline
\vspace{0.5cm}

{\bf Acknowledgements:} We thank CNPq (Brazil) for financial support.
The referees's remarks were important for the improvement of the final
version of this work. We thank also Jo\"el Tossa for his carefull reading
of the text.

\newpage
\centerline{\bf Figure captions}
\vspace{2.0cm}
\noindent
\vspace{0.5cm}
Figure 1: Behaviour of $h_1(\eta)$ for $n = 1$.\\
\vspace{0.5cm}
Figure 2: Behaviour of $h_2(\eta)$ for $n = 1$.\\
\vspace{0.5cm}
Figure 3: Behaviour of $h_3(\eta)$ for $n = 1$.\\
\vspace{0.5cm}
Figure 4: Behaviour of $h_4(\eta)$ for $n = 1$.\\
\vspace{0.5cm}
Figure 5: Behaviour of $\delta\phi(\eta)$ ($n = 0.5$) for the negative energy solutions.\\
\vspace{0.5cm}
Figure 6: Behaviour of $\delta\phi(\eta)$ for ($n = 0.5$) for the positive energy solutions.\\
Figure 7: Behaviour of $\Delta_M(\eta)$ for ($n = 0$) for the mixed model.\\

\end{document}